\documentclass[11pt]{article}
\pdfoutput=1

\usepackage[english]{babel}
\usepackage{amsmath,amssymb,mathrsfs,amsbsy,amstext}
\usepackage{graphicx}
\usepackage{amsfonts}
\usepackage{cite}

\begin{document}

\begin{titlepage}

\setcounter{page}{1} \baselineskip=15.5pt \thispagestyle{empty}

\bigskip\
\begin{center}
{\fontsize{17}{30}\selectfont  \bf Black hole formation in a contracting universe}
\end{center}

\vspace{0.5cm}
\begin{center}
{\fontsize{14}{30}\selectfont  Jerome Quintin,$^{*,}$\footnote{Vanier Canada Graduate Scholar}$^{,}$\footnote{Electronic address: jquintin@physics.mcgill.ca}
and Robert H.\ Brandenberger$^{*,}$\footnote{Electronic address: rhb@hep.physics.mcgill.ca}}
\end{center}

\begin{center}
\vskip 8pt
\textsl{${}^*$ Department of Physics, McGill University, Montr\'eal, QC, H3A 2T8, Canada}
\end{center}

\vspace{1.2cm}
\hrule \vspace{0.3cm}
{ \noindent \textbf{Abstract} \\[0.2cm]
\noindent
We study the evolution of cosmological perturbations in a contracting universe. We aim to determine under which conditions density perturbations grow to form large inhomogeneities and collapse into black holes. Our method consists in solving the cosmological perturbation equations in complete generality for a hydrodynamical fluid. We then describe the evolution of the fluctuations over the different length scales of interest and as a function of the equation of state for the fluid, and we explore two different types of initial conditions: quantum vacuum and thermal fluctuations. We also derive a general requirement for black hole collapse on sub-Hubble scales, and we use the Press-Schechter formalism to describe the black hole formation probability. For a fluid with a small sound speed (e.g., dust), we find that both quantum and thermal initial fluctuations grow in a contracting universe, and the largest inhomogeneities that first collapse into black holes are of Hubble size and the collapse occurs well before reaching the Planck scale. For a radiation-dominated fluid, we find that no black hole can form before reaching the Planck scale. In the context of matter bounce cosmology, it thus appears that only models in which a radiation-dominated era begins early in the cosmological evolution are robust against the formation of black holes. Yet, the formation of black holes might be an interesting feature for other models. We comment on a number of possible alternative early universe scenarios that could take advantage of this feature.}
 \vspace{0.3cm}
 \hrule

\vspace{0.6cm}

\end{titlepage}

\tableofcontents

\section{Introduction}\label{sec:intro}

The latest observations of the Cosmic Microwave Background
indicate that a good theoretical model for the very early universe
should predict a nearly scale-invariant power spectrum of curvature perturbations
with a small red tilt\ \cite{Ade:2015xua}, a small tensor-to-scalar ratio\ \cite{Array:2015xqh}, and small
non-Gaussianities\ \cite{Ade:2015ava}.
Inflationary cosmology\ \cite{Guth:1980zm,Mukhanov:1981xt,Linde:1981mu,Bardeen:1983qw} currently stands up as the best candidate for explaining
these observations\ \cite{Ade:2015lrj}.
Yet, it is still an incomplete theory conceptually\ \cite{Brandenberger:2010dk,Brandenberger:2011gk,Brandenberger:2012uj},
because, for example, it suffers from a singularity at the time of the Big Bang\ \cite{Borde:1993xh,Borde:2001nh}.
Thus, in addition to trying to resolve the issues of inflation,
it is helpful to study competitive or complementary ideas
that could enlighten our understanding of the very early universe.

One such idea is bouncing cosmology: one assumes that the universe existed forever before the Big Bang in a contracting phase,
after which it transitioned into the expending universe that we observe today.
In addition to solving the usual flatness and horizon problems of standard Big Bang cosmology,
assuming that quantum cosmological perturbations exit the Hubble horizon
in a matter-dominated contracting phase leads to a scale-invariant power spectrum of curvature perturbations\ \cite{Wands:1998yp,Finelli:2001sr}.
Furthermore, there exist many models that can avoid reaching a singularity at the time of the Big Bang,
hence leading to nonsingular bouncing cosmologies
(see\ \cite{Novello:2008ra,Brandenberger:2012zb,Cai:2014bea} and references therein).
Yet, it is still hard to construct models that can agree with all observational constraints
(see, e.g.,\ \cite{Quintin:2015rta} and also\ \cite{Battefeld:2014uga,Brandenberger:2016vhg} for reviews).

An additional difficulty with bouncing cosmology comes from the fact that it appears less robust against certain instabilities
as many unwanted features tend to grow in a contracting universe.
One example is anisotropies: as $a\rightarrow 0$, anisotropies grow as $\rho\propto a^{-6}$, whereas the background
matter and radiation evolve according to $\rho\propto a^{-3}$ and $\rho\propto a^{-4}$, respectively.
This is known as the Belinsky-Khalatnikov-Lifshitz (BKL) instability\ \cite{Belinsky:1970ew}.
This can be resolved if the background before the bounce can satisfy $\rho\propto a^{-q}$
with $q\gg 6$\ \cite{Erickson:2003zm,Cai:2013vm},
which naturally occurs within the Ekpyrotic model\ \cite{Khoury:2001wf,Khoury:2001zk} (see also\ \cite{Lehners:2008vx} and references therein).

There is another type of instability, always in a contracting universe, that has not been explored in as much detail,
namely the growth of inhomogeneities. This type of instability was already known from the 1960s\ \cite{Lifshitz&Khalatnikov63},
but it is only in the 2000s that the work was extended\ \cite{Banks:2002fe}, and it suggested that the growth of inhomogeneities
in a contracting universe could lead to the formation of black holes.

The goal of this paper is thus to revisit the analysis of the growth of inhomogeneities in a contracting universe,
and more specifically, characterize the formation of black holes. On one hand, we want to determine in which cases
a contracting universe is robust or not against the formation of large inhomogeneities and black holes. This will determine
in which cases it is justified to ignore the growth of inhomogeneities and allow us to claim which corresponding models remain healthy
or not. On the other hand, we want to determine in which cases a contracting universe inevitably leads to the formation of black holes.
These cases could be relevant in light of other alternative theories of the very early universe in which black holes
could be the seeds of the current universe.

The outline of this paper is as follows. First, in section\ \ref{sec:gravpot},
we begin by setting the general framework in which we work, and we solve for the evolution of the gravitational
potential in a contracting universe, aiming for generality. In section\ \ref{sec:densityJeansP}, we move on to find the density contrast
in a generic contracting universe, and we comment on its evolution over the different length scales of interest.
We also determine the power spectrum of the perturbations over the different scales of interest.
In section\ \ref{sec:ICs}, we explore two types of possible initial conditions for the fluctuations,
quantum vacuum initial conditions and thermal initial conditions,
and we find the power spectra in each cases. We also determine when the perturbations become non-linear.
Then, in section\ \ref{sec:BHformation}, we derive the condition for black hole collapse,
and we use the Press-Schechter formalism to determine which cases lead to the formation of black holes.
We also describe the black holes that form.
Finally, in section\ \ref{sec:discussion}, we summarize our results regarding the models that are robust (and those that are not)
against the formation of black holes. We end by suggesting possible alternative theories that could take advantage of the formation of
black holes.
Throughout this paper, we adopt the mostly minus convention for the metric,
and we define the reduced Planck mass by $M_\mathrm{Pl}\equiv(8\pi G_{\mathrm{N}})^{-1/2}$
where $G_{\mathrm{N}}$ is Newton's gravitational constant.

\section{Evolution of the gravitational potential in a contracting universe}\label{sec:gravpot}

\subsection{General background setup}

We begin by finding the general evolution of the cosmological perturbations in a contracting universe.
We try to be as generic as possible, and we do not specify any initial conditions for now.
We start with an action of the form
\begin{equation}
 S=-\frac{1}{16\pi G_{\mathrm{N}}}\int\mathrm{d}^4x~\sqrt{-g}R+S_\mathrm{m}~,
\end{equation}
where $g_{\mu\nu}$ is the metric tensor, $g\equiv\det(g_{\mu\nu})$,
$R$ is the Ricci scalar, and $S_\mathrm{m}$ is the action for matter.
We work in a flat Friedmann-Lema\^itre-Robertson-Walker (FLRW) universe, so the background metric is
\begin{equation}
 \mathrm{d}s^2=g_{\mu\nu}^{(0)}\mathrm{d}x^{\mu}\mathrm{d}x^{\nu}=a(\eta)^2(\mathrm{d}\eta^2-\delta_{ij}\mathrm{d}x^i\mathrm{d}x^j)~,
\end{equation}
where $a$ is the scale factor, $\eta$ is the conformal time (defined by $\mathrm{d}\eta\equiv a^{-1}\mathrm{d}t$, where $t$ is the physical time),
and the $x^i$'s represent the Cartesian comoving coordinates.
The energy-momentum tensor is defined by
\begin{equation}
 T_{\mu\nu}\equiv\frac{2}{\sqrt{-g}}\frac{\delta_gS_\mathrm{m}}{\delta g^{\mu\nu}}~,
\end{equation}
and we assume that it takes the form $T_\mu^{\ \nu}=\mathrm{diag}(\rho,-p\delta_i^{\ j})$,
where $p$ represents the pressure and $\rho$ the energy density.
Accordingly, the background equations of motion (EOMs) are
\begin{align}
\label{eq:Friedmann1}
 \mathcal{H}^2&=\frac{8\pi G_{\mathrm{N}}}{3}a^2\rho~, \\
\label{eq:Friedmann2}
 \mathcal{H}'&=-\frac{4\pi G_{\mathrm{N}}}{3}a^2\rho(1+3w)~,
\end{align}
where $'\equiv\mathrm{d}/\mathrm{d}\eta$ and $\mathcal{H}\equiv a'/a$ is the conformal Hubble parameter.
Furthermore, $w\equiv p/\rho$ is the equation of state (EoS) parameter.

From here on, we assume that the action for matter takes the form
\begin{equation}
 S_\mathrm{m}=-\int\mathrm{d}^4x~\sqrt{-g}\rho~,
\end{equation}
which is to say that \emph{we will work in a hydrodynamical fluid setup}.
The fluid has an EoS parameter $w$, and its sound speed is defined by
\begin{equation}
 c_{\mathrm{s}}^2\equiv\left(\frac{\partial p}{\partial\rho}\right)_s~,
\end{equation}
i.e.~it is the variation of the pressure with respect to the energy density
at constant entropy density, $s$. We note that we will ignore entropy perturbations throughout,
i.e.~\emph{we assume that the fluid has only adiabatic fluctuations}.

\subsection{Cosmological perturbations}\label{sec:cosmoperturbcontractinguniverse}

Let us introduce linear scalar perturbations about the background introduced above.
The perturbed metric written in the longitudinal (or conformal Newtonian) gauge with no anisotropic stress
(i.e.~$\delta T_{ij}=0$ for $i\neq j$) is
\begin{equation}
 \mathrm{d}s^2=a(\eta)^2\left\{\left[1+2\Phi(\eta,\mathbf{x})\right]\mathrm{d}\eta^2
 -\left[1-2\Phi(\eta,\mathbf{x})\right]\delta_{ij}\mathrm{d}x^i\mathrm{d}x^j\right\}~.
\end{equation}
The perturbation $\Phi$ is the Newtonian gravitational potential.
The resulting EOM from the perturbed Einstein equations gives rise to the following partial differential equation
\cite{Mukhanov:1990me},
\begin{equation}
\label{eq:EOMPhi}
 \Phi''+3\mathcal{H}(1+c_{\mathrm{s}}^2)\Phi'+[2\mathcal{H}'+(1+3c_{\mathrm{s}}^2)\mathcal{H}^2]\Phi-c_{\mathrm{s}}^2\nabla^2\Phi=0~,
\end{equation}
where $\nabla^2\equiv\partial_i\partial^i$ is the spacial Laplacian associated with the comoving space coordinates $x^i$.
Alternatively, using the Friedmann equations\ \eqref{eq:Friedmann1} and\ \eqref{eq:Friedmann2}
and transforming to Fourier space,
the EOM can be written as
\begin{equation}
 \Phi_k''+3\mathcal{H}(1+c_{\mathrm{s}}^2)\Phi_k'+3(c_{\mathrm{s}}^2-w)\mathcal{H}^2\Phi_k+c_{\mathrm{s}}^2k^2\Phi_k=0~,
\end{equation}
where $k$ represents the magnitude of the comoving wavenumber associated with the perturbations.

From here on, we will assume that we can split the cosmological evolution into one or more separate phases of constant
equation of state (EoS) parameter and constant sound speed.
Therefore, for a \emph{fixed (time-independent) EoS parameter} $w=\mathrm{constant}$, the solution to the background FLRW EOMs
is\footnote{Since we are interested in a contracting universe, we consider the physical time to be negative, i.e.~$t<0$.
The time $t=0$ would correspond to a possible Big Crunch, Big Bang, or bounce.
A negative physical time is equivalent to having a negative conformal time, $\eta<0$, when $w<-1$ or $w>-1/3$,
hence we have $(-\eta)$ in the scale factor since this quantity is positive.
We can safely restrict ourself to matter with $w>-1/3$ for the rest of this paper and ignore exotic matter
which could have $w<-1$. The case where $-1\leq w\leq -1/3$ should be analyzed separately, but
it will not be of interest in this paper.}
\begin{equation}
 a\propto(-\eta)^{\frac{2}{1+3w}}~,
\end{equation}
so
\begin{equation}
\label{eq:Hconformal}
 \mathcal{H}=-\frac{2}{1+3w}(-\eta)^{-1}~,
\end{equation}
and
\begin{equation}
\label{eq:Hconformalprime}
 \mathcal{H}'=-\frac{2}{1+3w}(-\eta)^{-2}~.
\end{equation}
The resulting EOM for the gravitational potential is
\begin{equation}
 \Phi_k''-\frac{6(1+c_\mathrm{s}^2)}{1+3w}\frac{1}{(-\eta)}\Phi_k'
 +\left(c_\mathrm{s}^2k^2+\frac{12(c_\mathrm{s}^2-w)}{(1+3w)^2}\frac{1}{(-\eta)^2}\right)\Phi_k=0~.
\end{equation}
For $w=\mathrm{constant}$ and for a \emph{fixed (time-independent) sound speed} $c_\mathrm{s}=\mathrm{constant}$,
the general solution to the above ordinary differential equation (ODE) is\footnote{We note that the above ODE is invariant
under $\eta \rightarrow -\eta$. Thus, the general solution is valid for both $\eta$ and $-\eta$.
We take the $-\eta$ branch of the solution for a contracting universe.}
\begin{equation}
\label{eq:gensolPhi}
 \Phi_k(\eta)=[2(1+3w)(-\eta)]^{\nu_1}\left[C_{1,k}J_{\nu_2}(-c_sk\eta)+C_{2,k}Y_{\nu_2}(-c_sk\eta)\right]~,
\end{equation}
where $C_{1,k}$ and $C_{2,k}$ are two constants of integration that will of set by the initial conditions.
Also, $J_\nu(x)$ and $Y_\nu(x)$ are the Bessel functions of the first and second kind, respectively.
Finally, for shorthand notation, we define the indices
\begin{equation}
\label{eq:nu1nu2}
 \nu_1\equiv-\frac{5+6c_\mathrm{s}^2-3w}{2(1+3w)}
 \qquad\mathrm{and}\qquad\nu_2\equiv\frac{\sqrt{25+12c_\mathrm{s}^2+36c_\mathrm{s}^4+18w-36c_\mathrm{s}^2w+9w^2}}{2(1+3w)}~.
\end{equation}
From the definition of $w$ and $c_\mathrm{s}$ and from the usual conservation equation [which follows from
equations\ \eqref{eq:Friedmann1} and\ \eqref{eq:Friedmann2}],
\begin{equation}
 \rho'+3\mathcal{H}\rho(1+w)=0~,
\end{equation}
it is straightforward to show that
\begin{equation}
 w'=3\mathcal{H}(1+w)(w-c_\mathrm{s}^2)~.
\end{equation}
Since a constant EoS parameter means $w'=0$, the above equation implies that $w=c_\mathrm{s}^2$
under our assumptions. In this situation, the indices defined above simplify to become
\begin{equation}
\label{eq:nu}
 \nu_1=-\frac{5+3w}{2(1+3w)}
 \qquad\mathrm{and}\qquad\nu_2=\frac{5+3w}{2(1+3w)}~,
\end{equation}
and hence, we define the index $\nu\equiv\nu_2=-\nu_1$ to simplify the notation from here on.

\section{Density contrast, Jeans scale, and power spectrum}\label{sec:densityJeansP}

\subsection{Density contrast}

The gauge-invariant density contrast in a flat universe, $\delta(\eta,\mathbf{x})$,
is related to the gravitation potential via\ \cite{Mukhanov:1990me}
\begin{equation}
 \delta\equiv\frac{\delta\rho^{(\mathrm{gi})}}{\rho^{(0)}}=\frac{2}{3\mathcal{H}^2}\left(\nabla^2\Phi-3\mathcal{H}\Phi'-3\mathcal{H}^2\Phi\right)~,
\end{equation}
where $\delta\rho(\eta,\mathbf{x})$ denotes the energy density fluctuations and $\rho^{(0)}(\eta)$ denotes the background energy density.
In Fourier space, this becomes
\begin{equation}
\label{eq:densitycontrastgeneral1}
 \delta_k\equiv\frac{\delta\rho^{(\mathrm{gi})}_k}{\rho^{(0)}}=-\frac{2}{3}\left(
 \frac{k^2}{\mathcal{H}^2}\Phi_k
 +\frac{3}{\mathcal{H}}\Phi_k'
 +3\Phi_k\right)~.
\end{equation}
Using equation\ \eqref{eq:Hconformal}, we have
\begin{equation}
\label{eq:densitycontrastgeneral2}
 \delta_k(\eta)=-\frac{(1+3w)^2}{6}k^2(-\eta)^2\Phi_k(\eta)+(1+3w)(-\eta)\Phi_k'(\eta)-2\Phi_k(\eta)~,
\end{equation}
and given the general solution for $\Phi_k(\eta)$, equation\ \eqref{eq:gensolPhi}, we get
\begin{align}
\label{eq:drhogen}
 \delta_k(\eta)=&~-\frac{1}{3\cdot 2^{\nu+1}(1+3w)^{\nu}(-\eta)^{\nu}}
 \Big\{6(1+3w)x[C_{1,k}J_{\nu-1}(x)+C_{2,k}Y_{\nu-1}(x)] \nonumber \\
 &+\Big[12-6(5+3w)+\frac{(1+3w)^2x^2}{c_\mathrm{s}^2}\Big]\Big[C_{1,k}J_{\nu}(x)+C_{2,k}Y_{\nu}(x)\Big]\Big\}~,
\end{align}
where we further define $x\equiv c_\mathrm{s}k(-\eta)$ for shorthand notation.

\subsection{Jeans scale}\label{sec:Jeanslength}

We will be interested in characterizing the formation of physical black holes, so we will primarily be interested in the sub-Hubble limit
of the above density contrast.
Since we are working with a fluid with a sound speed $c_\mathrm{s}$ possibly different from the speed of light,
there is another scale of interest, the Jeans scale.
It is defined to have a comoving wavenumber $k_\mathrm{J}$ such that the physical wavenumber is
\begin{equation}
 \frac{k_\mathrm{J}}{a}\equiv\frac{\sqrt{4\pi G_\mathrm{N}\rho}}{c_\mathrm{s}}=\sqrt{\frac{3}{2}}\frac{|H|}{c_\mathrm{s}}~,
\end{equation}
or alternatively, we can write
\begin{equation}
\label{eq:kJ}
 k_\mathrm{J}=\sqrt{\frac{3}{2}}\frac{|\mathcal{H}|}{c_\mathrm{s}}=\frac{\sqrt{6}}{1+3w}\frac{1}{c_\mathrm{s}|\eta|}~.
\end{equation}
The associated comoving wavelength is $\lambda_\mathrm{J}\equiv 2\pi/k_\mathrm{J}$.
Thus, the sub-Jeans scales correspond to the limit $\lambda\ll\lambda_\mathrm{J}$
or $k\gg k_\mathrm{J}$,
which is equivalent to the limit where $x$ is large;
the super-Jeans scales correspond to the limit $\lambda\gg\lambda_\mathrm{J}$
or $k\ll k_\mathrm{J}$,
which is equivalent to the limit where $x$ is small.
For dust, we have $w=c_\mathrm{s}^2\rightarrow 0$, and so $\lambda_\mathrm{J}\rightarrow 0$.
In other words, there is no sub-Jeans scales asymptotically, only super-Jeans/sub-Hubble and super-Hubble scales.
For radiation, we have $w=c_\mathrm{s}^2=1/3$, and so $k_\mathrm{J}=3/(\sqrt{2}|\eta|)$.
In comparison, the Hubble scale is given by
\begin{equation}
 k_H\equiv|\mathcal{H}|=\frac{2}{(1+3w)|\eta|}~,
\end{equation}
and so, for radiation, it is $k_H=1/|\eta|$. Thus, although we still have $\lambda_\mathrm{J}<\lambda_H$
in this case,
the two scales are really of the same order and nearly equal. Thus, there are very few scales in the super-Jeans/sub-Hubble regime.
Most scales are either sub-Jeans or super-Hubble in this case.

\subsubsection{Evolution below the Jeans length}\label{sec:subJeanslength}

We can expand the density contrast [equation\ \eqref{eq:drhogen}] to leading order in the limit where $x$ is large
($\lambda\ll\lambda_\mathrm{J}$) to find
\begin{align}
 \delta_k(\eta)\stackrel{\lambda\ll\lambda_\mathrm{J}}{\simeq}
 &-\frac{(1+3w)^{2-\nu}k^{3/2}(-\eta)^{3/2-\nu}}{6\cdot 2^{\nu}\sqrt{\pi c_\mathrm{s}}} \nonumber \\
\label{eq:drhoasym}
 &\times\left[(C_{1,k}-C_{2,k})\cos\left(c_\mathrm{s}k(-\eta)-\frac{\pi\nu}{2}\right)
 +(C_{1,k}+C_{2,k})\sin\left(c_\mathrm{s}k(-\eta)-\frac{\pi\nu}{2}\right)\right]~.
\end{align}
Thus, we see that the density contrast oscillates with frequency $\omega_k\equiv c_\mathrm{s}k$.
However, we are more interested in the amplitude which goes as $(-\eta)^{3/2-\nu}$. Recalling the definition of $\nu$ in
equation\ \eqref{eq:nu}, we note that
\begin{equation}
\label{eq:condnu1}
 \frac{3}{2}-\nu=\frac{3w-1}{3w+1}~.
\end{equation}
As physical time evolves in a contracting universe, $\eta\rightarrow 0^{-}$ or $(-\eta)\rightarrow 0^{+}$ for $w>-1/3$.
Thus, we see that the amplitude of the density contrast \emph{grows} in a contracting universe if
\begin{equation}
 \frac{3}{2}-\nu<0 \iff -\frac{1}{3}<w<\frac{1}{3}~;
\end{equation}
the amplitude of the density contrast is \emph{constant} if
\begin{equation}
 \frac{3}{2}-\nu=0 \iff w=\frac{1}{3}~;
\end{equation}
and the amplitude of the density contrast \emph{decreases} in a contracting universe if
\begin{equation}
 \frac{3}{2}-\nu>0 \iff w>\frac{1}{3}~.
\end{equation}
Consequently, in a dust-dominated contracting universe with $w=c_\mathrm{s}^2=0$,
we see that the amplitude of the density contrast \emph{grows} as $(-\eta)^{-1}$.
However, one needs to be careful since taking the limit $c_\mathrm{s}\rightarrow 0$ would also imply that the density contrast
blows up while the sub-Jeans regime of validity vanishes.
In fact, for normal baryonic matter (or even for dark matter), we expect the sound speed and the EoS parameter to be small,
but non-vanishing, i.e.~$0<w\ll 1$ and $0<c_\mathrm{s}\ll 1$.
For a radiation-dominated contracting universe ($w=c_\mathrm{s}^2=1/3$),
we find that the amplitude of the density contrast is constant.

\subsubsection{Evolution on super-Jeans/sub-Hubble scales}

On one hand, on super-Jeans scales, $x$ is small, and so, to leading order, equation\ \eqref{eq:gensolPhi} for the gravitational potential becomes
\begin{equation}
\label{eq:PhisuperJeansgeneral}
 \Phi_k(\eta)\simeq
 \frac{-C_{2,k}\Gamma(\nu)}{\pi(1+3w)^\nu c_\mathrm{s}^\nu k^\nu(-\eta)^{2\nu}}~,
\end{equation}
where $\Gamma(\nu)$ is the gamma function.
On the other hand, on super-Hubble scales, $k/\mathcal{H}$ is large in equation\ \eqref{eq:densitycontrastgeneral1},
and so, the density contrast reduces to
\begin{equation}
 \delta_k(\eta)\simeq-\frac{(1+3w)^2}{6}k^2(-\eta)^2\Phi_k(\eta)~.
\end{equation}
Therefore, substituting the super-Jeans solution for $\Phi_k$ into the sub-Hubble regime for $\delta_k$ yields
\begin{equation}
 \delta_k(\eta)\simeq\frac{(1+3w)^{2-\nu}\Gamma(\nu)}{6\pi c_\mathrm{s}^\nu}C_{2,k}k^{2-\nu}(-\eta)^{2(1-\nu)}~.
\end{equation}

The amplitude of the density contrast goes as $(-\eta)^{2(1-\nu)}$. Recalling the definition for $\nu$ is equation\ \eqref{eq:nu},
we note that
\begin{equation}
\label{eq:condnu2}
 2(1-\nu)=\frac{3(w-1)}{3w+1}~.
\end{equation}
Thus, we see that the amplitude of the density contrast \emph{grows} in a contracting universe if
\begin{equation}
 2(1-\nu)<0 \iff -\frac{1}{3}<w<1~;
\end{equation}
the amplitude of the density contrast is \emph{constant} if
\begin{equation}
 2(1-\nu)=0 \iff w=1~;
\end{equation}
and the amplitude of the density contrast \emph{decreases} in a contracting universe if
\begin{equation}
 2(1-\nu)>0 \iff w>1~.
\end{equation}
Consequently, in a dust-dominated and in a radiation-dominated contracting universe,
we find that the super-Jeans/sub-Hubble modes of the density contrast grow in amplitude
as they approach a possible bounce.

\subsubsection{Evolution on super-Hubble scales}

On super-Hubble scales, the general form for the density contrast is $\delta_k(\eta)\simeq-2(\Phi_k'/\mathcal{H}+\Phi_k)$.
Substituting in equation\ \eqref{eq:PhisuperJeansgeneral}, the super-Hubble (and necessarily super-Jeans) solution for
the density contrast is
\begin{equation}
 \delta_k(\eta)\simeq\frac{2[1-\nu(1+3w)]\Gamma(\nu)}{\pi (1+3w)^\nu c_\mathrm{s}^\nu}C_{2,k}k^{-\nu}(-\eta)^{-2\nu}~.
\end{equation}
This time, the amplitude goes as $(-\eta)^{-2\nu}$, but $-2\nu<0$ given our assumption that $w>-1/3$,
indicating growth in the amplitude of the perturbations in all cases.

\subsection{Power spectrum}

We saw above that the resulting density contrast oscillates with a time-varying amplitude on sub-Jeans scales. Since we will be primarily interested
in the amplitude, let us average out the oscillations. Moreover, the general solution in Fourier space is generally complex,
so let us take the magnitude squared to get the more physically meaningful real amplitude. Thus, equation\ \eqref{eq:drhoasym} becomes
\begin{equation}
\label{eq:aveP1}
 \left\langle\left|\delta_k(\eta)\right|^2\right\rangle\stackrel{\lambda\ll\lambda_\mathrm{J}}{\simeq}
 \frac{(1+3w)^{2(2-\nu)}}{36\cdot 2^{2\nu}\pi c_\mathrm{s}}(-\eta)^{3-2\nu}k^{3}
 \left(\left|C_{1,k}\right|^2+\left|C_{2,k}\right|^2\right)~.
\end{equation}
Here, $\langle\cdot\rangle$ really means averaging over the oscillations,
i.e.~$\langle\cos[\omega_k(-\eta)-\pi\nu/2]\sin[\omega_k(-\eta)-\pi\nu/2]\rangle=0$
and $\langle\cos^2[\omega_k(-\eta)-\pi\nu/2]\rangle=\langle\sin^2[\omega_k(-\eta)-\pi\nu/2]\rangle=1/2$.

On super-Jeans/sub-Hubble scales, we simply have
\begin{equation}
\label{eq:deltaksquaredsuperJeanssubHubblegeneral}
 \left|\delta_k(\eta)\right|^2\simeq\frac{(1+3w)^{2(2-\nu)}\Gamma(\nu)^2}{36\pi^2 c_\mathrm{s}^{2\nu}}
 |C_{2,k}|^2k^{2(2-\nu)}(-\eta)^{4(1-\nu)}~,
\end{equation}
and on super-Hubble scales, it is
\begin{equation}
 \left|\delta_k(\eta)\right|^2\simeq\frac{4[1-\nu(1+3w)]^2\Gamma(\nu)^2}{\pi^2 (1+3w)^{2\nu} c_\mathrm{s}^{2\nu}}
 |C_{2,k}|^2k^{-2\nu}(-\eta)^{-4\nu}~.
\end{equation}

One can interpret the above quantities as the power spectra of the density contrast,
i.e.~$P_\delta(k,\eta)\equiv|\delta_k(\eta)|^2$. In dimensionless form,
\begin{equation}
 \mathcal{P}_\delta(k,\eta)\equiv\frac{k^3}{2\pi^2}P_\delta(k,\eta)=\frac{k^3}{2\pi^2}|\delta_k(\eta)|^2~.
\end{equation}
The averaged power spectrum on sub-Jeans scales is then identified with equation\ \eqref{eq:aveP1}, so denoting the average by a bar, we have
\begin{equation}
\label{eq:PdeltageneralsubJeans}
 \bar{\mathcal{P}}_{\delta}(k,\eta)
 =\frac{k^3}{2\pi^2}\left\langle\left|\delta_k(\eta)\right|^2\right\rangle
 \stackrel{\lambda\ll\lambda_\mathrm{J}}{\simeq}
 \frac{(1+3w)^{2(2-\nu)}}{72\cdot 2^{2\nu}\pi^3c_\mathrm{s}}(-\eta)^{3-2\nu}k^{6}
 \left(\left|C_{1,k}\right|^2+\left|C_{2,k}\right|^2\right)~.
\end{equation}
Equivalently, on super-Jeans/sub-Hubble scales, the power spectrum is
\begin{equation}
\label{eq:PdeltageneralsuperJeanssubHubble}
 \mathcal{P}_{\delta}(k,\eta)
 =\frac{k^3}{2\pi^2}\left|\delta_k(\eta)\right|^2
 \simeq\frac{(1+3w)^{2(2-\nu)}\Gamma(\nu)^2}{72\pi^4 c_\mathrm{s}^{2\nu}}
 |C_{2,k}|^2k^{7-2\nu}(-\eta)^{4(1-\nu)}~,
\end{equation}
and on super-Hubble scales, we have
\begin{equation}
 \mathcal{P}_{\delta}(k,\eta)\simeq\frac{2[1-\nu(1+3w)]^2\Gamma(\nu)^2}{\pi^4 (1+3w)^{2\nu} c_\mathrm{s}^{2\nu}}
 |C_{2,k}|^2k^{3-2\nu}(-\eta)^{-4\nu}~.
\end{equation}

\section{Examples of initial conditions}\label{sec:ICs}

\subsection{Quantum vacuum}\label{sec:BDICs}

At this point, we did not specify the initial conditions, which is why the above resulting power spectra still depend on the
integration constants $C_{1,k}$ and $C_{2,k}$. A typical initial condition would be a quantum (Bunch-Davies) vacuum,
\begin{equation}
\label{eq:vkini}
 v_k^{(\mathrm{ini})}(\eta)=\frac{\mathrm{e}^{-\mathrm{i}c_\mathrm{s}k\eta}}{\sqrt{2c_\mathrm{s}k}}~,\qquad c_\mathrm{s}k(-\eta)\rightarrow\infty~,
\end{equation}
where $v$ is the Mukhanov-Sasaki variable.
For example, in matter bounce cosmology, matching the super-Hubble evolution of the cosmological perturbations
in a matter-dominated contracting universe at Hubble crossing with quantum vacuum initial conditions yields
a scale-invariant power spectrum of curvature perturbations\ \cite{Wands:1998yp,Finelli:2001sr}.
Initially, one has an oscillating quantum vacuum state, but at Hubble radius crossing,
the quantum fluctuations `squeeze' and emerge as classical fluctuations on super-Hubble scales
(see, e.g.,\ \cite{Polarski:1995jg,Battarra:2013cha,Pinto-Neto:2013npa,Stargen:2016cft}).
Similarly, in the case where the Jeans length is different than the Hubble radius, we associate quantum vacuum
fluctuations with the sub-Jeans regime, and the squeezing of the fluctuations at Jeans length crossing leads to
classical perturbations on super-Jeans scales, the growth of which might lead to the formation of black holes.
Since we showed above that the Jeans and Hubble lengths are nearly equal for radiation,
the quantum vacuum will be of more interest for dust when $w=c_\mathrm{s}^2\ll 1$.
Yet, we must ensure that we do not set $c_\mathrm{s}=0$,
since otherwise, the Jeans length would vanish, and we would no longer be able to define a quantum vacuum state on sub-Jeans scales.

We need to relate the variable $v$ with the density contrast for which we computed the power spectrum.
The Mukhanov-Sasaki variable is related to the gravitational potential via\ \cite{Mukhanov:1990me}
\begin{equation}
 \nabla^2\Phi=-\frac{\beta}{\sqrt{2}M_{\mathrm{Pl}}c_{\mathrm{s}}^2\mathcal{H}}\left(\frac{v}{z}\right)'~,
\end{equation}
where
\begin{equation}
 z\equiv\frac{a\sqrt{\beta}}{c_\mathrm{s}\mathcal{H}}~,
\end{equation}
and
\begin{equation}
 \beta\equiv\mathcal{H}^2-\mathcal{H}'~.
\end{equation}
Upon transforming to Fourier space, the initial condition in terms of the gravitational potential reads
\begin{equation}
 -k^2\Phi_k^{(\mathrm{ini})}
 =-\frac{\beta}{\sqrt{2}M_{\mathrm{Pl}}c_{\mathrm{s}}^2\mathcal{H}}\left(\frac{v_k^{(\mathrm{ini})}}{z}\right)'~.
\end{equation}
Using equation\ \eqref{eq:vkini}, we have $v_k^{(\mathrm{ini})\prime}=-\mathrm{i}c_\mathrm{s}kv_k^{(\mathrm{ini})}$, and so,
the above becomes
\begin{equation}
 \Phi_k^{(\mathrm{ini})}=-\frac{\beta v_k^{(\mathrm{ini})}}{\sqrt{2}M_\mathrm{Pl}c_\mathrm{s}^2\mathcal{H}k^2z}\left(\mathrm{i}c_\mathrm{s}k+\frac{z'}{z}\right)
 =-\frac{\mathrm{i}\beta(-\eta)^{3/2}}{2M_\mathrm{Pl}\mathcal{H}z}\frac{\mathrm{e}^{\mathrm{i}x}}{x^{3/2}}\left(1-\frac{z'}{z}\frac{\mathrm{i}(-\eta)}{x}\right)~.
\end{equation}
where again, $x\equiv c_\mathrm{s}k(-\eta)$. To leading order when $x$ is large, i.e.~on sub-Jeans scales,
the initial condition for $\Phi_k$ becomes
\begin{equation}
\label{eq:ICPhi}
 \Phi_k^{(\mathrm{ini})}\stackrel{x\rightarrow\infty}{\simeq}-\frac{\mathrm{i}\beta(-\eta)^{3/2}}{2M_\mathrm{Pl}\mathcal{H}z}\frac{\mathrm{e}^{\mathrm{i}x}}{x^{3/2}}~.
\end{equation}

We recall that we found the general solution for $\Phi_k(\eta)$ in equation\ \eqref{eq:gensolPhi}.
We now demand that the large $x$ limit of equation\ \eqref{eq:gensolPhi} matches the above expression for $\Phi_k^{(\mathrm{ini})}$.
To leading order, equation\ \eqref{eq:gensolPhi} becomes
\begin{equation}
 \Phi_k(\eta)\stackrel{x\rightarrow\infty}{\simeq}
 \frac{\mathrm{e}^{\mathrm{i}(x-\frac{\nu\pi}{2}-\frac{\pi}{4})}(C_{1,k}-\mathrm{i}C_{2,k})
 +\mathrm{e}^{-\mathrm{i}(x-\frac{\nu\pi}{2}-\frac{\pi}{4})}(C_{1,k}+\mathrm{i}C_{2,k})}
 {2^{\nu+\frac{1}{2}}\sqrt{\pi}(1+3w)^{\nu}(-\eta)^{\nu}x^{1/2}}~.
\end{equation}
For the above to match with the initial condition [equation\ \eqref{eq:ICPhi}], which only goes as
$\mathrm{e}^{\mathrm{i}x}$, it is clear that we must have $C_{1,k}+\mathrm{i}C_{2,k}=0$,
so that the term $\mathrm{e}^{-\mathrm{i}x}$ goes to $0$ in the above.
Thus, $C_{1,k}=-\mathrm{i}C_{2,k}$, and the above becomes
\begin{equation}
 \Phi_k(\eta)\stackrel{x\rightarrow\infty}{\simeq}
 -\mathrm{i}\frac{2^{\frac{1}{2}-\nu}\mathrm{e}^{\mathrm{i}(x-\frac{\nu\pi}{2}-\frac{\pi}{4})}}{\sqrt{\pi}(1+3w)^{\nu}(-\eta)^{\nu}x^{1/2}}C_{2,k}~.
\end{equation}
Equating this to equation\ \eqref{eq:ICPhi} imposes
\begin{align}
 C_{2,k}&=\frac{2^{\nu-\frac{3}{2}}\sqrt{\pi}(1+3w)^{\nu}(-\eta)^{\nu+\frac{1}{2}}\beta}{c_\mathrm{s}M_\mathrm{Pl}\mathcal{H}zk}
 \mathrm{e}^{\mathrm{i}(\nu+\frac{1}{2})\frac{\pi}{2}} \nonumber \\
 &=\frac{2^{\nu-\frac{3}{2}}\sqrt{\pi}(1+3w)^{\nu}(-\eta)^{\nu+\frac{1}{2}}\sqrt{\mathcal{H}^2-\mathcal{H}'}}{M_\mathrm{Pl}ak}
 \mathrm{e}^{\mathrm{i}(\nu+\frac{1}{2})\frac{\pi}{2}}~,
\end{align}
where we use the definition of $\beta$ and $z$ to simplify the second equality.
Using equations\ \eqref{eq:Hconformal} and\ \eqref{eq:Hconformalprime} for $\mathcal{H}$ and $\mathcal{H}'$,
the integrating constant further simplifies to become
\begin{equation}
 C_{2,k}=\frac{2^{\nu-\frac{3}{2}}\sqrt{6\pi(1+w)}(1+3w)^{\nu-1}(-\eta)^{\nu-\frac{1}{2}}}{M_\mathrm{Pl}ak}
 \mathrm{e}^{\mathrm{i}(\nu+\frac{1}{2})\frac{\pi}{2}}~.
\end{equation}
The left-over time-dependent factor is actually just a constant after simplification since
\begin{equation}
 \frac{(-\eta)^{\nu-\frac{1}{2}}}{a(\eta)}=\frac{\eta_0^{\frac{2}{1+3w}}}{a_0}
\end{equation}
if we normalize the scale factor as $a(\eta)=a_0(-\eta/\eta_0)^{\frac{2}{1+3w}}$.
However, it will be more convenient to keep the scale factor in the expression.
In the end, we are left with
\begin{equation}
\label{eq:C2k2}
 |C_{2,k}|^2=\frac{3\pi(1+w)(1+3w)^{2(\nu-1)}(-\eta)^{2\nu-1}}{2^{2(1-\nu)}M_\mathrm{Pl}^2a^2k^2}~.
\end{equation}

Substituting the above integration constant found for quantum vacuum initial conditions on sub-Jeans scales
into the general power spectrum on super-Jeans/sub-Hubble scales [equation\ \eqref{eq:PdeltageneralsuperJeanssubHubble}] leads to
\begin{equation}
\label{eq:PdeltageneralsuperJeanssubHubbleBDICs}
 \mathcal{P}_{\delta}(k,\eta)
 \simeq\frac{2^{2\nu}(1+w)(1+3w)^{2}\Gamma(\nu)^2}{96\pi^3 c_\mathrm{s}^{2\nu}M_\mathrm{Pl}^2a^2}
 k^{5-2\nu}(-\eta)^{3-2\nu}~.
\end{equation}
As we saw earlier, the super-Jeans/sub-Hubble regime is valid for dust, but not so much for radiation.
Thus, to leading order when $w=c_\mathrm{s}^2\ll 1$, we obtain
\begin{equation}
 \mathcal{P}_{\delta}(k,\eta)
 \simeq\frac{3}{16\pi^2 c_\mathrm{s}^{5}M_\mathrm{Pl}^2a^2}(-\eta)^{-2}
 =\frac{3\mathcal{H}^2}{64\pi^2 c_\mathrm{s}^{5}M_\mathrm{Pl}^2a^2}~,
\end{equation}
or
\begin{equation}
\label{eq:Pdeltadust}
 \mathcal{P}_{\delta}(k,t)\simeq\frac{3H^2(t)}{64\pi^2 c_\mathrm{s}^{5}M_\mathrm{Pl}^2}~,
\end{equation}
where we use the fact that $\mathcal{H}=aH$ with $H\equiv\mathrm{d}\ln a/\mathrm{d}t$ being the physical Hubble parameter.
As a result, the density contrast power spectrum on super-Jeans/sub-Hubble scales is scale-invariant (independent of $k$)
and grows in amplitude as time evolves ($|H|$ grows in time in a contracting universe).

We will soon be interested in describing the formation of black holes. A necessary condition (but not sufficient) for black
hole formation is $\mathcal{P}_\delta(k,t)>1$, which can be viewed as defining the scale of non-linearity.
However, since we obtain a scale-invariant power spectrum,
there is no specific non-linear scale, but rather, a non-linear time after which all super-Jeans/sub-Hubble modes become non-linear.
In terms of the Hubble parameter, non-linearity is reached when
\begin{equation}
\label{eq:NLtimemattersuperJeans}
 |H|\gtrsim\frac{8}{\sqrt{3}}\pi c_\mathrm{s}^{5/2}M_\mathrm{Pl}~.
\end{equation}
Hence, all super-Jeans/sub-Hubble scales become non-linear
when the energy scale of the universe becomes larger than a fraction of the Planck scale.
The fraction may be very small depending on the smallness of the sound speed, and so,
this may occur well before a Planck time before a possible bounce.

\subsection{Thermal initial conditions}\label{sec:TVICs}

As another example of initial conditions, let us consider the case of thermal fluctuations.
In this situation, the averaged energy density fluctuations on sub-Jeans scales for a thermal statistical system of characteristic size $L$
and temperature $T$ is given by (see\ \cite{Magueijo:2002pg,Cai:2009rd,Biswas:2013lna} and also\ \cite{Peebles:1994xt})
\begin{equation}
 \langle\delta^2\rangle_L=\frac{T^2}{L^3\rho^2}\frac{\partial\rho}{\partial T}~.
\end{equation}
In Fourier space, the averaged density fluctuations become
\begin{equation}
\label{eq:deltakthermalgeneral}
 |\delta_k|^2=\frac{\gamma_\mathrm{f}^2T^2}{a^3\rho^2}\frac{\partial\rho}{\partial T}~,
\end{equation}
where the constant $\gamma_\mathrm{f}$ depends on the choice of window function when doing the Fourier transformation (see\ \cite{Biswas:2013lna}
for details).

In our context of an ideal fluid with EoS parameter $w$, one can express the energy density as a function of temperature by
(see\ \cite{Biswas:2013lna})
\begin{equation}
\label{eq:rhoofT}
 \rho(T)=\frac{m_T^4}{w}\left(\frac{T}{m_T}\right)^{\frac{1+w}{w}}~,
\end{equation}
where $m_T$ is a preferred mass scale associated with the fluid.
We note that the above expression is only valid for $0<w<1$, so when we consider dust, we will take the limit $0<w\ll 1$ as before.
For radiation ($w=1/3$), we recover the Stefan-Boltzmann law $\rho(T)\propto T^4$, where there is no preferred mass scale.
For a general EoS, taking $\rho(a)=\rho_0(a/a_0)^{-3(1+w)}$
and using the above expression for $\rho(T)$, equation\ \eqref{eq:deltakthermalgeneral} becomes
\begin{equation}
\label{eq:deltaksquaredsubJeansgeneral}
 |\delta_k|^2=\frac{\gamma_\mathrm{f}^2(1+w)}{a_0^3m_T^3}\left(\frac{m_T^4}{w\rho_0}\right)^{\frac{1}{1+w}}~.
\end{equation}
Setting $w=1/3$ for radiation, the resulting dimensionless power spectrum on sub-Jeans scales is
\begin{equation}
\label{eq:PdeltaradiationthermalICs}
 \mathcal{P}_\delta(k)=\frac{2\gamma_\mathrm{f}^2}{3^{1/4}\pi^2}\frac{1}{\rho_0^{3/4}}\left(\frac{a}{a_0}\right)^{3}\left(\frac{k}{a}\right)^3
 =\frac{2\gamma_\mathrm{f}^2}{3\pi^2M_\mathrm{Pl}^{3/2}|H|^{3/2}}\left(\frac{k}{a}\right)^3~,
\end{equation}
where in the second equality, we use the Friedmann equation $3M_\mathrm{Pl}^2H^2=\rho$ and equation\ \eqref{eq:rhoofT} (the Stefan-Boltzmann law
for radiation).
We note that the power spectrum is blue, and also, it is time independent for a fixed comoving wavenumber $k$.
It follows that the scales that are non-linear ($\mathcal{P}_\delta>1$) must satisfy
\begin{equation}
 \frac{k}{a}>\left(\frac{3}{2}\right)^{1/3}\left(\frac{\pi}{\gamma_\mathrm{f}}\right)^{2/3}\sqrt{M_\mathrm{Pl}|H|}~.
\end{equation}
Thus, at later times, when the energy scale $|H|$ of the universe is higher,
there are fewer physical scales $k/a$ that become non-linear.
However, as $|H|\rightarrow 0$ in the infinite past, it would appear that all physical scales become non-linear,
which seems to render unphysical this choice of initial conditions.
Yet, there is a subtlety that allows us to still consider thermal initial conditions.

We note that thermal fluctuations can be interpreted as a Poisson process, which presupposes a set of regions
with coherence length $\ell_C$ (see\ \cite{Magueijo:2002pg}). In fact, we can express the averaged density contrast in position space as
\begin{equation}
 \langle\delta^2\rangle_L=\left(\frac{\ell_C}{L}\right)^3~,
\end{equation}
where the temperature-dependent coherence length is given by $\ell_C^3=(T/\rho)^2(\partial\rho/\partial T)$.
For example, for radiation, we find that $\ell_C\propto T^{-1}\propto a$.
With this interpretation, a requirement for the fluctuations to be non-linear
is that the coherence length must be larger than the scale of the thermal system, i.e.~$\ell_C>L$.
However, a requirement for thermalization is that $\ell_C\ll L$, and so, when the thermal fluctuations become non-linear,
it must follow that they are no more thermal.
In particular, this implies that in the far past when $\ell_C$ is large, one cannot consider thermal fluctuations
on arbitrary small length scales.
In this sense, the thermal initial state is well defined in the far past as long as we consider length scales that are large enough
compared to the coherence length at that time. These set the initial conditions, and as the fluctuations evolve
gravitationally without interactions, they loose their thermality, and in particular, they may become non-linear.

For dust in the limit $w\ll 1$, the density contrast squared, equation\ \eqref{eq:deltaksquaredsubJeansgeneral}, should go to
\begin{equation}
 |\delta_k|^2\simeq\frac{\gamma_\mathrm{f}^2m_T}{w\rho_0a_0^3}
\end{equation}
on sub-Jeans scales.
In comparison, the general solution for dust on sub-Jeans scales, equation\ \eqref{eq:aveP1} with $w=c_\mathrm{s}^2\ll 1$, is
\begin{equation}
 \langle|\delta_k|^2\rangle\simeq\frac{k^3}{1152\pi c_\mathrm{s}(-\eta)^2}\left(|C_{1,k}|^2+|C_{2,k}|^2\right)~.
\end{equation}
Thus, if the $C_{1,k}$ and $C_{2,k}$ terms contribute equally, it follows that we must take
\begin{equation}
\label{eq:C2k2thermaldust}
 |C_{2,k}|^2=\frac{576\pi\gamma_\mathrm{f}^2c_\mathrm{s}m_T(-\eta_\mathrm{ini})^2}{w\rho_0a_0^3k^3}~,
\end{equation}
where $\eta_\mathrm{ini}$ is the initial conformal time at which the initial conditions are set.
In other words, at $\eta_\mathrm{ini}$,
we set the hydrodynamical cosmological perturbations to have the amplitude and spectrum of thermal fluctuations.
From that moment onward, and especially as we consider the super-Jeans regime, the fluctuations are no more thermal.

On super-Jeans/sub-Hubble scales, the power spectrum for dust (equation\ \eqref{eq:deltaksquaredsuperJeanssubHubblegeneral} with $w=0$)
is
\begin{equation}
 |\delta_k(\eta)|^2\simeq\frac{|C_{2,k}|^2}{64\pi c_\mathrm{s}^5k(-\eta)^6}~.
\end{equation}
Substituting in equation\ \eqref{eq:C2k2thermaldust} yields
\begin{equation}
\label{eq:PdeltadustthermalICs}
 \mathcal{P}_\delta(k,t)\simeq\frac{3\gamma_\mathrm{f}^2}{32\pi^2c_\mathrm{s}^4w}\left(\frac{a_\mathrm{ini}}{a}\right)
 \left(\frac{H}{M_\mathrm{Pl}}\right)^2m_T\left(\frac{a}{k}\right)~,
\end{equation}
where we use the relations $\rho=\rho_0(a/a_0)^{-3}=3M_\mathrm{Pl}^2H^2$,
$(-\eta)=-2/\mathcal{H}=-2/(aH)$, and $a/a_\mathrm{ini}=(\eta/\eta_\mathrm{ini})^2$ to simplify the expression,
and we note that we define $a_\mathrm{ini}\equiv a(\eta_\mathrm{ini})$.
It follows that non-linearity occurs when
\begin{equation}
 \frac{k}{a}<\frac{3\gamma_\mathrm{f}^2}{32\pi^2c_\mathrm{s}^4w}\left(\frac{a_\mathrm{ini}}{a}\right)
 \left(\frac{H}{M_\mathrm{Pl}}\right)^2m_T~.
\end{equation}
Thus, as $|H|/M_\mathrm{Pl}$ and $a_\mathrm{ini}/a$ grow in a contracting universe,
more physical scales become non-linear on super-Jeans/sub-Hubble scales.
Also, it appears that the largest length scales become non-linear first.
This suggests that larger black holes form before
smaller ones. We will confirm this result in the following section.

\section{Black hole formation}\label{sec:BHformation}

The evolution and the spectrum of the cosmological perturbations found in the previous sections
allow us to address the question of black hole formation.
We found that the amplitude of the perturbations increases in many instances, and so, we expect some of the overdensities
to collapse to form black holes as one approaches a possible bounce.
However, we need to determine under what conditions one can claim that a black has formed.

\subsection{General requirement for black hole collapse}

Let us consider an element of physical volume $\mathrm{d}V=\mathrm{d}^3\mathbf{q}$ at some physical time $t$ and physical position $\mathbf{q}$.
Then, the amount of mass excess enclosed in this physical volume element at position $\mathbf{q}$ as a function of time is given by
\begin{equation}
\label{eq:dM}
 \mathrm{d}\delta M(t,\mathbf{q})=\mathrm{d}^3\mathbf{q}~\delta\rho(t,\mathbf{q})~.
\end{equation}
We argue that a black hole forms when an amount of mass excess $\delta M\geq M_s$ is found inside a ball of radius $R\leq R_s$,
where $R_s$ is the (physical) Schwarzschild radius
given by $R_s=2M_sG_\mathrm{N}$ for a black hole of mass
$M_s$. This appears to be a fair requirement assuming that the hoop conjecture holds (the original idea
of the hoop conjecture comes from\ \cite{Thorne,Misner:1974qy} ;
see also subsequent papers on the subject, e.g.~\cite{Flanagan91,Gibbons:2009xm}). Therefore, a black hole forms,
i.e.~an event horizon appears, if
\begin{equation}
 \int_{R\leq R_s}\mathrm{d}\delta M\geq M_s~.
\end{equation}
More precisely, using equation\ \eqref{eq:dM} for the mass excess element, we say that the condition
for a black hole to form at the point $\mathbf{q}_{\star}$ and time $t_{\star}$ is
\begin{equation}
\label{eq:conditiongeneral}
 \int_{\mathcal{B}(R_s,\mathbf{q}_{\star})}\mathrm{d}^3\mathbf{q}~\delta\rho(t_{\star},\mathbf{q})\geq\frac{R_s}{2G_\mathrm{N}}~,
\end{equation}
where the integral is over the volume of a ball of radius $R_s$ centered at $\mathbf{q}_{\star}$,
or more formally, over the region
\begin{equation}
 \mathcal{B}(R_s,\mathbf{q}_{\star})\equiv\left\{\mathbf{q}\in\mathbb{R}^3\Big||\mathbf{q}-\mathbf{q}_{\star}|\leq R_s\right\}~.
\end{equation}

\subsection{Smoothing}

The goal is thus to evaluate the integral on the left-hand side of equation\ \eqref{eq:conditiongeneral}.
In order to do so, let us review the idea of smoothing.
In general, the definition of a smoothed perturbation $\delta$ over a characteristic scale $\mathscr{R}$ is
\begin{equation}
 \delta(t,\mathbf{x};\mathscr{R})\equiv\frac{1}{\mathscr{V}(\mathscr{R})}\int\mathrm{d}^3\mathbf{\tilde x}
 ~W(|\mathbf{x}-\mathbf{\tilde x}|/\mathscr{R})\delta(t,\mathbf{\tilde x})~,
\end{equation}
where $W$ is the window function, assumed to be spherically symmetric with comoving radius $\mathscr{R}$.
The comoving volume associated with the smoothing region of characteristic scale $\mathscr{R}$ is defined by
\begin{equation}
 \mathscr{V}(\mathscr{R})\equiv\int\mathrm{d}^3\mathbf{x}~W(|\mathbf{x}|/\mathscr{R})=4\pi\mathscr{R}^3\int\mathrm{d}y~y^2W(y)~,
\end{equation}
where $\mathbf{y}\equiv\mathbf{x}/\mathscr{R}$, $y\equiv|\mathbf{y}|$.
Then, in Fourier space, $\delta_{\mathbf{k}}(t;\mathscr{R})=\mathcal{W}(k\mathscr{R})\delta_{\mathbf{k}}(t)$,
where we denote the Fourier transform of $W$ by $\mathcal{W}$.
Also, the variance is related to the power spectrum by
\begin{equation}
\label{eq:variancegeneral}
 \sigma^2(\mathscr{R},t)\equiv\langle[\delta(t,\mathbf{x};\mathscr{R})]^2\rangle
 =\int_0^\infty\frac{\mathrm{d}k}{k}~\mathcal{W}^2(k\mathscr{R})\mathcal{P}_{\delta}(k,t)
 =\frac{1}{2\pi^2}\int_0^\infty\mathrm{d}k~k^2\mathcal{W}^2(k\mathscr{R})|\delta_{k}(t)|^2~.
\end{equation}

A commonly used window function is the top-hat window function,
\begin{equation}
 W(y)=\begin{cases}
       1 & \qquad\mathrm{for}\ 0\leq y\leq 1\ (|\mathbf{x}|\leq \mathscr{R})\\
       0 & \qquad\mathrm{for}\ y>1\ (|\mathbf{x}|>\mathscr{R})~.
      \end{cases}
\end{equation}
Its Fourier transform is
\begin{equation}
\label{eq:tildeWtophat}
 \mathcal{W}(k\mathscr{R})=\frac{3[\sin(k\mathscr{R})-k\mathscr{R}\cos(k\mathscr{R})]}{(k\mathscr{R})^3}~.
\end{equation}
Using the top-hat window function, we recognize that smoothing the perturbation $\delta\rho$ with
characteristic scale\footnote{$R_s$ is the physical Schwarzschild radius, so we divide by the scale factor to have
a comoving quantity.} $R_s/a$ yields
\begin{align}
 \delta\rho(t,\mathbf{x};R_s/a)&=\frac{1}{\mathscr{V}(R_s/a)}\int\mathrm{d}^3\mathbf{\tilde x}
 ~W(a|\mathbf{x}-\mathbf{\tilde x}|/R_s)\delta\rho(t,\mathbf{\tilde x}) \\
 &=\frac{3a^3}{4\pi R_s^3}\int_{|\mathbf{x}-\mathbf{\tilde x}|\leq\frac{R_s}{a}}\mathrm{d}^3\mathbf{\tilde x}~\delta\rho(t,\mathbf{\tilde x})~,
\end{align}
but since comoving coordinates $\mathbf{x}$ are related to physical coordinates $\mathbf{q}$ by $\mathbf{q}=a\mathbf{x}$, we get
\begin{equation}
 \delta\rho(t,\mathbf{x};R_s/a)
 =\frac{3}{4\pi R_s^3}\int_{|\mathbf{q}-\mathbf{\tilde q}|\leq R_s}\mathrm{d}^3\mathbf{\tilde q}~\delta\rho(t,\mathbf{\tilde q})
 \equiv\delta\rho(t,\mathbf{q};R_s)~.
\end{equation}
Therefore, we notice that the left-hand side of equation\ \eqref{eq:conditiongeneral}
is simply related to the smoothed perturbation $\delta\rho(t,\mathbf{q};R_s)$.
Specifically, at a fixed time and position, we find
\begin{equation}
 \int_{\mathbf{\tilde q}\in\mathcal{B}(R_s,\mathbf{q}_\star)}\mathrm{d}^3\mathbf{\tilde q}~\delta\rho(t_\star,\mathbf{\tilde q})
 =\frac{4\pi}{3}R_s^3\delta\rho(t_\star,\mathbf{q}_\star;R_s)~.
 \label{eq:integraldeltarhodV}
\end{equation}

\subsection{Critical density contrast for black hole collapse}

Combining equations\ \eqref{eq:conditiongeneral} and \eqref{eq:integraldeltarhodV}, we say that
a black hole forms when
\begin{equation}
 \delta\rho(t_{\star},\mathbf{q}_{\star};R_s)\geq\frac{3}{8\pi G_\mathrm{N}R_s^2}~.
\end{equation}
Dividing by the background energy density on both sides and recalling the Friedmann
equation $H(t)^2=8\pi G_\mathrm{N}\rho^{(0)}(t)/3$, the condition becomes
\begin{equation}
 \frac{\delta\rho}{\rho^{(0)}}(t_{\star},\mathbf{q}_{\star};R_s)\geq\left(\frac{H^{-1}(t_{\star})}{R_s}\right)^2~.
\end{equation}
We define the critical density contrast as a function of physical size $R$ and time $t$ to be
\begin{equation}
\label{eq:deltacphysical}
 \delta_\mathrm{c}(R,t)\equiv\left(\frac{H^{-1}(t)}{R}\right)^2~.
\end{equation}
Alternatively, as a function of conformal time and for a comoving scale $\mathscr{R}=R/a$,
\begin{equation}
\label{eq:deltacconformal}
 \delta_\mathrm{c}(\mathscr{R},\eta)=\left(\frac{\mathcal{H}^{-1}(\eta)}{\mathscr{R}}\right)^2~.
\end{equation}
Finally, the condition to form a black hole of Schwarzschild radius $R_s$ at any time $t$ and position $\mathbf{q}$ is
\begin{equation}
 \delta(t,\mathbf{q};R_s)\geq\delta_\mathrm{c}(R_s,t)~;
\end{equation}
or at conformal time $\eta$ and comoving position $\mathbf{x}$, a black hole forms if
$\delta(\eta,\mathbf{x};\mathscr{R}_s)\geq\delta_\mathrm{c}(\mathscr{R}_s,\eta)$,
where $\mathscr{R}_s\equiv R_s/a$ is the comoving Schwarzschild radius.

From the form of our critical density contrast equation\ \eqref{eq:deltacphysical}, we notice that a necessary (but not sufficient) condition
for black hole formation is $\delta>1$ on scales where $R<|H|^{-1}$, i.e.~on sub-Hubble scales.
This is to be expected since black holes are highly non-linear objects.
In general, the smaller $R$ is compared to the Hubble radius, then the larger $\delta_\mathrm{c}$ is,
and so, the larger the density contrast $\delta$ needs to be to form a black hole of size $R$.
In other words, the smaller the black hole we want to form, the more difficult it becomes.
However, since $\delta$ is a smoothed quantity in the condition $\delta>\delta_\mathrm{c}$,
i.e.~integrated over space, its particular spectrum will affect the condition for black hole formation.
For example, more power on smaller scales could lead to the production of smaller black holes before
larger black holes.
This is why we focused on computing the spectrum of $\delta$ in Fourier space in the previous sections.

We point out that equation\ \eqref{eq:deltacphysical} is only valid for $R\leq|H|^{-1}$. Naively, it is obvious that this equation
cannot hold for super-Hubble perturbations since for long wavelength fluctuations, the critical density contrast for black hole formation
would become very small, which would imply that small fluctuations would collapse into large black holes. In fact, if one takes
$R\rightarrow\infty$, then $\delta_\mathrm{c}\rightarrow 0$, and it would seem to imply that \emph{any} fluctuation would collapse into a black hole,
which is physically inadmissible. The underlying reason comes from the fact that no black hole horizon can actually form above
the cosmological apparent horizon. Indeed, any observer inside the cosmological horizon cannot know about the existence of the formation of a black
hole if this black hole's horizon is greater than the cosmological horizon. Yet, there can still be large density fluctuations
on super-Hubble scales, and these can form black holes if they re-enter the cosmological horizon at later times.
In the context of a nonsingular bounce\footnote{However, in the context of a nonsingular bounce,
one would need to consider the possible effect of the formation of sub-Hubble black holes,
so it is not yet clear how density fluctuations would evolve through a nonsingular bounce.},
large density fluctuations that exit the Hubble radius in the contracting phase
could collapse into black holes once they re-enter the Hubble radius in the expanding phase.
This is similar to the formation of primordial black holes\ \cite{Carr:1974nx,Carr:1975qj} in inflation where large density fluctuations
can exit the Hubble radius during the inflationary phase and re-enter the Hubble radius in the subsequent radiation-dominated
expanding phase, at which point the large density fluctuations can collapse into black holes
(see, e.g.,\ \cite{GarciaBellido:1996qt,Yokoyama:1999xi}).
The fact that no black hole horizon can form on length scales larger than the cosmological horizon is also
explicit in general relativistic constructions such as in Schwarzschild-de Sitter spacetime
(see, e.g.,\ \cite{Bousso:2002fq,Shankaranarayanan:2003ya,Faraoni:2015ula} and references therein)
or McVittie spacetime (see, e.g.,\ \cite{Kaloper:2010ec,Faraoni:2012gz,Faraoni:2015ula} and references therein).

\subsection{Press-Schechter formalism and a condition for black hole collapse}
Following the idea of the Press-Schechter formalism\ \cite{Press:1973iz},
we say that $\delta(\eta,\mathbf{x})$ is a Gaussian random field, and thus, the fraction of mass in spheres of radius $\mathscr{R}$
with overdensity $\delta>\delta_\mathrm{c}$ has a Gaussian probability,
\begin{equation}
\label{eq:probcollapse}
 \mathscr{P}(\mathscr{R},\eta)=\frac{1}{\sqrt{2\pi}\sigma(\mathscr{R},\eta)}
 \int_{\delta_\mathrm{c}(\mathscr{R},\eta)}^{\infty}\mathrm{d}\delta~\exp\left[-\frac{\delta^2}{2\sigma^2(\mathscr{R},\eta)}\right]
 =\frac{1}{2}~\mathrm{erfc}\left[\frac{\delta_\mathrm{c}(\mathscr{R},\eta)}{\sqrt{2}\sigma(\mathscr{R},\eta)}\right]~,
\end{equation}
where we recall that the variance $\sigma^2(\mathscr{R},\eta)$ is given by equation\ \eqref{eq:variancegeneral}
(simply replacing physical time with conformal time in this case).
To account for the fact that there is an equal amount of matter in underdense as in overdense regions, relative to the background,
we say that the actual probability is
\begin{equation}
\label{eq:trueprobcollapse}
 F(\mathscr{R},\eta)=2\mathscr{P}(\mathscr{R},\eta)=\mathrm{erfc}\left[\frac{\delta_\mathrm{c}(\mathscr{R},\eta)}{\sqrt{2}\sigma(\mathscr{R},\eta)}\right]~.
\end{equation}
Accordingly, the probability to form a black hole of comoving size $\mathscr{R}$ at conformal time $\eta$
is large when the ratio $\delta_\mathrm{c}(\mathscr{R},\eta)/\sigma(\mathscr{R},\eta)$ is small.
In fact, $F\rightarrow 1$ as $\delta_\mathrm{c}/\sigma\rightarrow 0$.
Therefore, it is fair to say that black holes of characteristic radius $\mathscr{R}$ can only form in significant numbers
when $\sigma(\mathscr{R},\eta)\gtrsim\delta_\mathrm{c}(\mathscr{R},\eta)$
(see, e.g., \cite{Moetal}).
This makes sense intuitively since we found earlier that a black hole was formed at position $\mathbf{x}$ when
$\delta(\eta,\mathbf{x};\mathscr{R})\geq\delta_\mathrm{c}(\mathscr{R},\eta)$. Now, we say that a necessary condition
is $\sigma(\mathscr{R},\eta)=\sqrt{\langle[\delta(\eta,\mathbf{x};\mathscr{R})]^2\rangle}\gtrsim\delta_\mathrm{c}(\mathscr{R},\eta)$,
which is more or less equivalent.

In general, we evaluate $\sigma^2$ as follows:
\begin{equation}
\label{eq:sigma2generalPdelta}
 \sigma^2(\mathscr{R},\eta)=\int_0^\infty\frac{\mathrm{d}k}{k}\mathcal{W}^2(k\mathscr{R})\mathcal{P}_\delta(k,\eta)
 \simeq\int_{k_H}^{k_\mathrm{J}}\frac{\mathrm{d}k}{k}\mathcal{W}^2(k\mathscr{R})\mathcal{P}_\delta(k,\eta)
 +\int_{k_\mathrm{J}}^\infty\frac{\mathrm{d}k}{k}\mathcal{W}^2(k\mathscr{R})\mathcal{P}_\delta(k,\eta)~.
\end{equation}
We note that instead of integrating from $k=0$, we set an infrared cutoff at the Hubble scale $k_H=|\mathcal{H}|=a|H|$
since we argue that no black holes could form on super-Hubble scales.
In general, on super-Jeans/sub-Hubble scales,
$\mathcal{P}_\delta(k,\eta)$ is given by equation\ \eqref{eq:PdeltageneralsuperJeanssubHubble},
and on sub-Jeans scales, $\mathcal{P}_\delta(k,\eta)$ is given by equation\ \eqref{eq:PdeltageneralsubJeans}.
Accordingly, one could determine the general expression for the variance on arbitrary scales and for arbitrary matter,
but the two most interesting cases, dust and radiation, are only applicable on distinct scales.
Thus, we put generality aside, and we only consider them separately below.

\subsubsection{Dust on super-Jeans/sub-Hubble scales}

Let us begin with dust with quantum vacuum initial conditions. In this case, the density contrast power spectrum
on super-Jeans/sub-Hubble scales is given by equation\ \eqref{eq:Pdeltadust}, and so, the variance
is found to be
\begin{equation}
 \sigma^2(\mathscr{R},t)=\int_{k_H}^{k_\mathrm{J}}\frac{\mathrm{d}k}{k}~\mathcal{W}^2(k\mathscr{R})\mathcal{P}_\delta(k,t)
 \simeq\frac{3H^2}{64\pi^2 c_\mathrm{s}^{5}M_\mathrm{Pl}^2}
 \int_{k_H}^{k_\mathrm{J}}\mathrm{d}k~\frac{\mathcal{W}^2(k\mathscr{R})}{k}~.
\end{equation}
Taking equation\ \eqref{eq:tildeWtophat} for the top-hat window function, the integral reduces to
\begin{equation}
 \int_{k_H}^{k_\mathrm{J}}\mathrm{d}k~\frac{\mathcal{W}^2(k\mathscr{R})}{k}
 =\frac{7}{4}-\gamma-\ln(2k_H\mathscr{R})+\frac{(k_H\mathscr{R})^2}{10}+\mathcal{O}[(k_H\mathscr{R})^4]+\mathcal{O}[(k_\mathrm{J}\mathscr{R})^{-4}]~,
\end{equation}
where we use the fact that $k_\mathrm{J}\mathscr{R}\gg 1$ on super-Jeans scales and $k_H\mathscr{R}\ll 1$ on sub-Hubble scales.
In the above, $\gamma\approx 0.577$ is the Euler-Mascheroni constant, which appears in the series expansion of the cosine integral.
Keeping only the constant and logarithmic terms to leading order, the variance is found to be
\begin{equation}
\label{eq:sigma2dustquantumICs}
 \sigma^2(R,t)\simeq\frac{3H^2}{64\pi^2 c_\mathrm{s}^{5}M_\mathrm{Pl}^2}\left[\frac{7}{4}-\gamma-\ln(2|H|R)\right]~,
\end{equation}
where we use $k_H\mathscr{R}=|\mathcal{H}|\mathscr{R}=a|H|\mathscr{R}=|H|R$.
Then, recalling equation\ \eqref{eq:deltacphysical}, the condition for black hole formation, $\sigma\gtrsim\delta_\mathrm{c}$, reads
\begin{equation}
 \frac{3H^6R^4}{64\pi^2 c_\mathrm{s}^{5}M_\mathrm{Pl}^2}\left[\frac{7}{4}-\gamma-\ln 2-\frac{1}{2}\ln(H^2R^2)\right]\gtrsim 1~.
\end{equation}
This expression cannot be reduced analytically, so let us consider the formation of black holes which have a radius equal to a fraction of the Hubble
radius, i.e.~let $R=\alpha|H|^{-1}$ for some constant $\alpha\leq 1$ not too small so that we remain on super-Jeans scales.
In this case, the above condition for black hole formation reduces to
\begin{equation}
 |H|\gtrsim\frac{8\pi c_\mathrm{s}^{5/2}M_\mathrm{Pl}}{\sqrt{3}\alpha^2}\left[\frac{7}{4}-\gamma-\ln 2-\ln(\alpha)\right]^{-1/2}~.
\end{equation}
We see that the larger $\alpha$ is, the smaller the expression on the right-hand side of the above condition,
which implies that a smaller energy scale (smaller $|H|$) needs to be reached to form black holes of size $R=\alpha|H|^{-1}$.
In particular, this implies that Hubble-size black holes, i.e.~black holes with Schwarzschild radius $R=|H|^{-1}$,
form first when
\begin{equation}
 |H|\simeq\frac{8\pi c_\mathrm{s}^{5/2}M_\mathrm{Pl}}{\sqrt{3(7/4-\gamma-\ln 2)}}~,
\end{equation}
a small fraction of the Planck scale when $c_\mathrm{s}$ is small.
In comparison, we found in equation\ \eqref{eq:NLtimemattersuperJeans} that we entered the non-linear regime when
$|H|\simeq 8\pi c_\mathrm{s}^{5/2}M_\mathrm{Pl}/\sqrt{3}$.

For dust with thermal initial conditions, the density contrast power spectrum
on super-Jeans/sub-Hubble scales is given by equation\ \eqref{eq:PdeltadustthermalICs}, and so, the variance
is found to be
\begin{equation}
 \sigma^2(\mathscr{R},t)\simeq\frac{3\gamma_\mathrm{f}^2a_\mathrm{ini}m_T}{32\pi^2c_\mathrm{s}^4w}
 \left(\frac{H}{M_\mathrm{Pl}}\right)^2
 \int_{k_H}^{k_\mathrm{J}}\mathrm{d}k~\frac{\mathcal{W}^2(k\mathscr{R})}{k^2}~.
\end{equation}
Using the top-hat window function again, the integral reduces to
\begin{equation}
 \int_{k_H}^{k_\mathrm{J}}\mathrm{d}k\frac{\mathcal{W}^2(k\mathscr{R})}{k^2}\simeq\mathscr{R}\left\{-\frac{9\pi}{35}
 +\mathcal{O}[(k_\mathrm{J}\mathscr{R})^{-5}]+\frac{1}{k_H\mathscr{R}}+\frac{k_H\mathscr{R}}{5}-\frac{(k_H\mathscr{R})^3}{175}
 +\mathcal{O}[(k_H\mathscr{R})^5]\right\}~,
\end{equation}
in the limits $k_H\mathscr{R}\ll 1$ and $k_\mathrm{J}\mathscr{R}\gg 1$. Keeping the leading order terms and converting to physical quantities,
the variance reduces to
\begin{equation}
\label{eq:sigma2dustthermalICs}
 \sigma^2(R,t)\simeq\frac{3\gamma_\mathrm{f}^2}{32\pi^2c_\mathrm{s}^4w}\left(\frac{a_\mathrm{ini}}{a}\right)m_T\left(\frac{H}{M_\mathrm{Pl}}\right)^2
 \left(\frac{1}{|H|}-\frac{9\pi R}{35}\right)~.
\end{equation}
The condition for black hole formation $\sigma\gtrsim\delta_\mathrm{c}$ becomes
\begin{equation}
 \frac{3\gamma_\mathrm{f}^2}{32\pi^2c_\mathrm{s}^4w}\left(\frac{a_\mathrm{ini}}{a}\right)\frac{m_TH^6R^4}{M_\mathrm{Pl}^2}
 \left(\frac{1}{|H|}-\frac{9\pi R}{35}\right)\gtrsim 1~.
\end{equation}
As before, let us consider black holes with radius $R=\alpha|H|^{-1}$, i.e.~a fraction $\alpha\leq 1$ of the Hubble radius.
Also, we note that, since $\rho\propto a^{-3}$ and $\rho\propto H^2$, we have $a/a_\mathrm{ini}=(H_\mathrm{ini}/H)^{2/3}$.
Thus, the condition reduces to
\begin{equation}
 |H|\gtrsim\frac{8}{3^{3/5}}\left(\frac{\pi}{\gamma_\mathrm{f}}\right)^{6/5}c_\mathrm{s}^{12/5}w^{3/5}\alpha^{-12/5}
 \left(1-\frac{9\pi\alpha}{35}\right)^{-3/5}\frac{H_\mathrm{ini}^{2/5}M_\mathrm{Pl}^{6/5}}{m_T^{3/5}}~.
\end{equation}
We see that the larger the fraction $\alpha$ is, the earlier black holes form.
Consequently, Hubble-size black holes form first once again in this case.
Associating the preferred mass scale of the fluid $m_T$ with the energy scale at the time at which the initial conditions
are taken, i.e.~letting $m_T=H_\mathrm{ini}$, we find that
\begin{equation}
 |H|\simeq 8\left(\frac{\pi}{\gamma_\mathrm{f}}\right)^{6/5}\left[3\left(1-\frac{9\pi}{35}\right)\right]^{-3/5}c_\mathrm{s}^{12/5}w^{3/5}
 \left(\frac{M_\mathrm{Pl}}{H_\mathrm{ini}}\right)^{1/5}M_\mathrm{Pl}
\end{equation}
corresponds to the Hubble parameter at the time that the first (Hubble-size) black holes are formed.
On one hand, since $c_\mathrm{s}^{12/5}w^{3/5}=c_\mathrm{s}^{18/5}\ll 1$ for dust, the critical time for black hole
formation may be well before a Planck time before a possible bounce. On the other hand,
we normally consider initial conditions such that $H_\mathrm{ini}\ll M_\mathrm{Pl}$,
so this pushes the critical time closer to the Planck time.

\begin{figure*}
\centering
\includegraphics[scale=0.75]{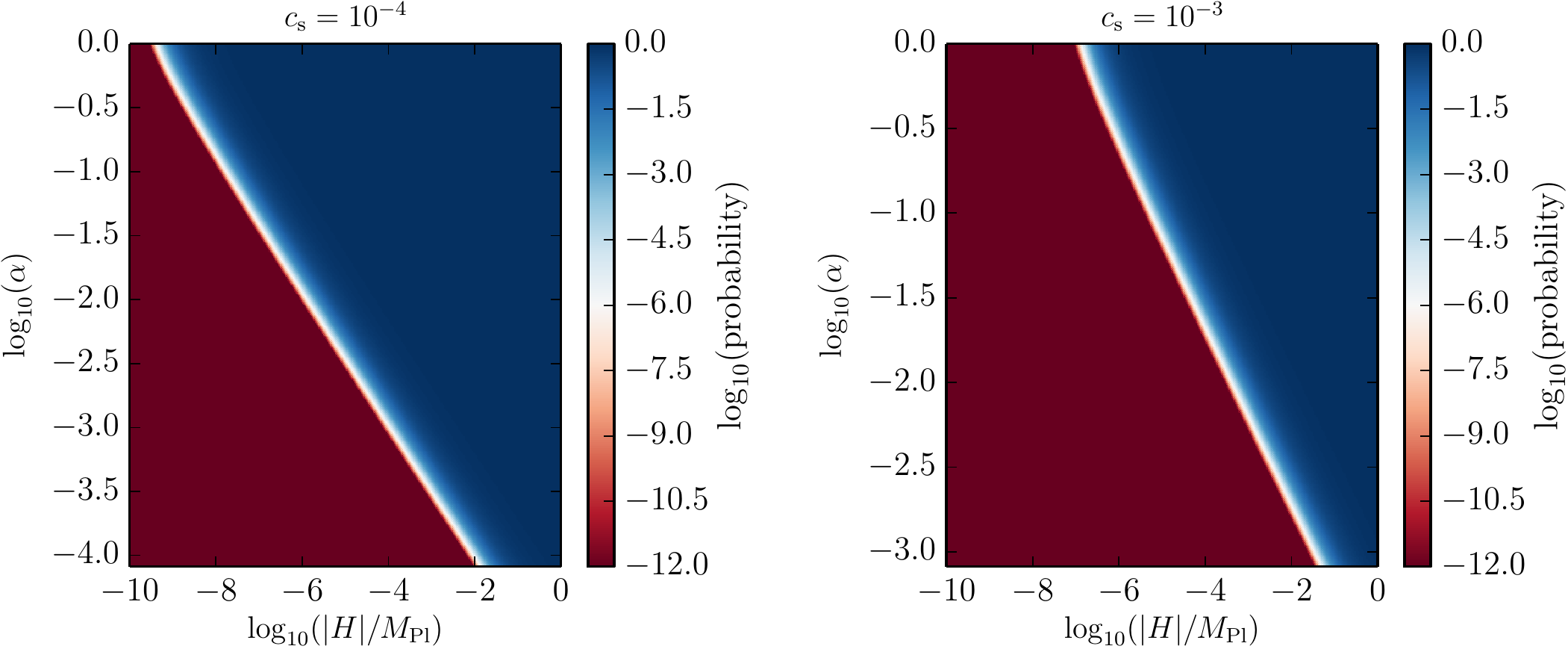}
\caption{Plots of the probability $F=\mathrm{erfc}[\delta_\mathrm{c}/(\sqrt{2}\sigma)]$
that a black hole has formed with dimensionless radius $\alpha=R/|H|^{-1}$ (vertical axis) and
at Hubble parameter $|H|$ (horizontal axis). The probability is color coded on a $\log_{10}$ scale.
The left and right plots show the probability for a sound speed $c_\mathrm{s}$ of $10^{-4}$
and $10^{-3}$, respectively.
The form of $\sigma$ is taken from equation\ \eqref{eq:sigma2dustquantumICs} for quantum vacuum initial conditions.}
\label{fig1}
\end{figure*}

\begin{figure*}
\centering
\includegraphics[scale=0.75]{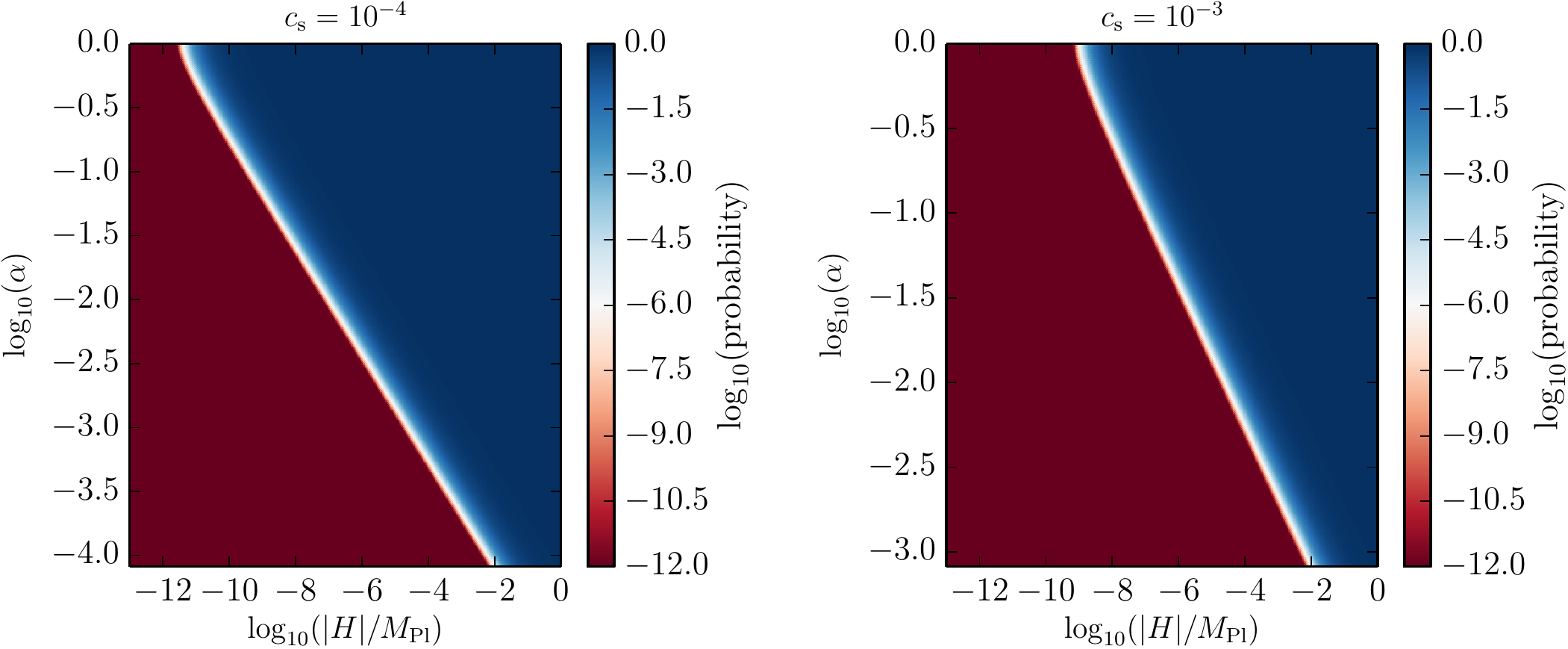}
\caption{Same plots as in figure\ \ref{fig1}, but for thermal initial conditions,
so the form of $\sigma$ is taken from equation\ \eqref{eq:sigma2dustthermalICs}.
Also, we take $H_\mathrm{ini}=10^{-16}M_\mathrm{Pl}$ and $\gamma_\mathrm{f}=2\sqrt{2}\pi^{3/4}$.}
\label{fig2}
\end{figure*}

To visualize the above results, we plot the probability of black hole formation,
equation\ \eqref{eq:trueprobcollapse}, in figures\ \ref{fig1} and\ \ref{fig2} for different cases.
The plots are color coded on a $\log_{10}$ scale
in terms of the probability $F$ ranging from $1$ ($100\%$ chance that a black hole has formed ; in dark blue)
to\footnote{We actually put a hard cutoff at $\log_{10}F=-12$, because as the probability goes to $0$,
the log scale would go to $-\infty$. As $F\leq 10^{-12}$, the probability is negligible, and we associate it with $0$
probability.} $0$ (no chance that a black hole has formed ; in dark red).
In each plot, the vertical axis represents the radius of the black hole as a fraction of the Hubble radius,
i.e.~$\alpha=R/|H|^{-1}$, on a $\log_{10}$ scale, ranging from the Jeans scale $\alpha=a|H|/k_\mathrm{J}=\sqrt{2/3}c_\mathrm{s}$
to the Hubble scale $\alpha=1$. The horizontal axis represents the Hubble parameter as a fraction of the Planck scale on a $\log_{10}$ scale.

In figure\ \ref{fig1}, we show the probability in the case of quantum vacuum initial conditions,
so we take equation\ \eqref{eq:sigma2dustquantumICs} for $\sigma^2$. The left plot shows the result with a sound speed of $c_\mathrm{s}=10^{-4}$
and the right plot shows $c_\mathrm{s}=10^{-3}$. In all cases, we see that the probability to form a black hole changes abruptly from
nearly $0$ (dark red) to nearly $1$ (dark blue).
In the left plot, the lowest energy scale at which this occurs is around $|H|\sim 10^{-9}M_\mathrm{Pl}$,
at which point $\alpha\sim 1$, meaning that the first black holes that form are of Hubble size. With a larger sound speed,
in the right plot, the same is true, but it occurs when the Hubble parameter is $|H|\sim 10^{-7}M_\mathrm{Pl}$.
In other words, the first black holes would form later with a larger sound speed.

In figure\ \ref{fig2}, we show the probability in the case of thermal initial conditions,
so we take equation\ \eqref{eq:sigma2dustthermalICs} for $\sigma^2$. In addition, we pick the Hubble parameter
at the initial time to be $H_\mathrm{ini}=10^{-16}M_\mathrm{Pl}$, and following\ \cite{Biswas:2013lna},
we take $\gamma_\mathrm{f}=2\sqrt{2}\pi^{3/4}$.
Again, the left and right plots show $c_\mathrm{s}=10^{-4}$ and $c_\mathrm{s}=10^{-3}$, respectively.
Just like in figure\ \ref{fig1}, the transition from a low probability of finding black holes to a high probability is abrupt,
and the plots show that Hubble-size black holes ($\alpha=1$) form first. The critical value of the Hubble parameter
is quantitatively different but remains well below the Planck scale for a small sound speed.

\subsubsection{Radiation on sub-Jeans scales}

For radiation with thermal initial conditions, the density contrast power spectrum
on sub-Jeans scales is given by equation\ \eqref{eq:PdeltaradiationthermalICs}, and so, the variance
is found to be
\begin{equation}
 \sigma^2(\mathscr{R},t)\simeq\frac{2\gamma_\mathrm{f}^2}{3\pi^2a^3M_\mathrm{Pl}^{3/2}|H|^{3/2}}
 \int_{k_\mathrm{J}}^\infty\mathrm{d}k~\mathcal{W}^2(k\mathscr{R})k^2~.
\end{equation}
Taking the top-hat window function, equation\ \eqref{eq:tildeWtophat}, the integral reduces to
\begin{equation}
 \int_{k_\mathrm{J}}^\infty\mathrm{d}k~\mathcal{W}^2(k\mathscr{R})k^2\simeq\mathscr{R}^{-3}
 \left\{\frac{3\pi}{2}-\frac{1}{3}(k_\mathrm{J}\mathscr{R})^3+\mathcal{O}[(k_\mathrm{J}\mathscr{R})^5]\right\}
\end{equation}
in the sub-Jeans limit $k_\mathrm{J}\mathscr{R}\ll 1$.
Keeping only the leading order term, the variance becomes
\begin{equation}
 \sigma^2(R,t)\simeq\frac{\gamma_\mathrm{f}^2}{\pi M_\mathrm{Pl}^{3/2}|H|^{3/2}R^3}~.
\end{equation}
Therefore, the condition for black hole formation $\sigma\gtrsim\delta_\mathrm{c}$ reduces to
\begin{equation}
 |H|\gtrsim\left(\frac{\pi^2M_\mathrm{Pl}^3}{\gamma_\mathrm{f}^4R^2}\right)^{1/5}~.
\end{equation}
Thus, the larger $R$ is, the smaller the quantity on the right-hand of the above expression, and the earlier
black hole formation occurs. Since this is only valid on sub-Jeans scales, it implies that Jeans-size black holes
form first. Taking $R=(a/k_\mathrm{J})=\sqrt{2}/(3|H|)$ for radiation, these black holes form when
\begin{equation}
 |H|\simeq\left(\frac{9\pi^2}{2\gamma_\mathrm{f}^4}\right)^{1/3}M_\mathrm{Pl}~.
\end{equation}
The numerical constant $[9\pi^2/(2\gamma_\mathrm{f}^4)]^{1/3}$ is only of\footnote{For example, it is shown in\ \cite{Biswas:2013lna}
that a Gaussian window function yields $\gamma_\mathrm{f}=2\sqrt{2}\pi^{3/4}$, and so,
$[9\pi^2/(2\gamma_\mathrm{f}^4)]^{1/3}\approx 0.28$, which is marginally smaller than $\mathcal{O}(1)$.} $\mathcal{O}(1)$, and consequently,
the first black holes in a radiation-dominated contracting universe with thermal initial conditions form only
when the energy scale reaches the Planck scale.

In nonsingular bouncing cosmologies, one usual assumes that new physics
appears in the effective theory well below the Planck scale, and thus, one could expect to enter the bounce without forming
any black hole. In this sense, a nonsingular bouncing cosmology in which the radiation-dominated contracting phase
starts early in its cosmological evolution is robust against the formation of black holes.
However, it remains to be shown that such an early transition from matter domination to radiation domination
does not spoil the scale invariance of the power spectrum of curvature perturbations at the scales of observational interest
in the usual matter bounce.

\section{Conclusions and discussion}\label{sec:discussion}

In this paper, we studied the adiabatic cosmological perturbations of a hydrodynamical fluid with constant EoS parameter and
constant sound speed in a flat\footnote{Although we did not include the possible effects
of spatial curvature in our analysis, we believe that our results would not be greatly affected by those effects
since the contribution of spatial curvature decreases in a contracting universe.}
contracting universe. We found the general evolution of the density contrast over the different
regimes of interest: sub-Jeans scales, super-Jeans/sub-Hubble scales, and super-Hubble scales. The key results
that are independent of the initial conditions can be summarized as follows:
\begin{itemize}
 \item for a radiation-dominated contracting universe, the amplitude of the density contrast on sub-Jeans scales
 is constant in time;
 \item for a matter-dominated contracting universe, the amplitude of the density contrast on super-Jeans/sub-Hubble scales
 grows with time as one approaches a possible bounce.
\end{itemize}
We then considered two sets of initial conditions: quantum vacuum initial conditions and thermal initial conditions.
This allowed us to find the general form of the power spectrum, and the main results are given below.
\begin{itemize}
 \item By setting quantum vacuum initial conditions at Jeans crossing, the density contrast power spectrum
 in a matter-dominated contracting universe on super-Jeans/sub-Hubble scales is scale invariant,
 grows as $H^2(t)/M_\mathrm{Pl}^2$, and is enhanced by the smallness of the sound speed ($\mathcal{P}_\delta\sim c_\mathrm{s}^{-5}$).
 In addition, we find that non-linearity is reached well before the Planck scale when the sound speed is small.
 \item By setting thermal initial conditions at a fixed time on sub-Jeans scales, the density contrast power spectrum
 in a radiation-dominated contracting universe (on sub-Jeans scales) is blue ($\mathcal{P}_\delta\sim k^3$),
 and, in a matter-dominated contracting universe (on super-Jeans/sub-Hubble scales), it is red ($\mathcal{P}_\delta\sim k^{-1}$).
 Accordingly, for radiation, non-linearity occurs first on smaller length scales (Planck scale), whereas for matter, non-linearity
 occurs first on larger length scales (Hubble scale).
\end{itemize}
Then, under the assumption that the hoop conjecture is valid, we derived a general requirement for black hole collapse.
By smoothing out the density contrast power spectrum and using the Press-Schechter formalism to describe the probability
of black hole formation, we arrived at the following final results.
\begin{itemize}
 \item For a matter-dominated contracting universe with quantum vacuum initial conditions, Hubble-size black holes,
 i.e.~black holes with Schwarzschild radius $R=|H|^{-1}$, form first when the Hubble parameter reaches
 $|H|\sim c_\mathrm{s}^{5/2}M_\mathrm{Pl}$, a small fraction of the Planck scale for $c_\mathrm{s}\ll 1$.
 \item We find the same results when we take thermal initial conditions instead of a quantum state,
 except that the critical energy scale for black hole formation goes as
 $|H|\sim c_\mathrm{s}^{18/5}(M_\mathrm{Pl}/H_\mathrm{ini})^{1/5}M_\mathrm{Pl}$,
 which depends on the value of the Hubble parameter at the time that the initial conditions are taken.
 Yet, in most cases, this is still a small fraction of the Planck scale.
 \item For a radiation-dominated contracting universe with thermal initial conditions,
 no black hole can form before the Hubble parameter reaches $|H|\simeq [9\pi^2/(2\gamma_\mathrm{f}^4)]^{1/3}M_\mathrm{Pl}\sim M_\mathrm{Pl}$,
 i.e.~order the Planck scale.
\end{itemize}

In light of these results, we showed in this paper that nonsingular bouncing cosmology is robust against the formation of black holes
if the sound speed is large enough. In particular, for a radiation-dominated contracting universe with $c_\mathrm{s}^2=1/3$,
we found that no black hole could form before reaching a Planck time before the bounce. Equivalently, we expect this result to hold
for even stiffer equations of state. In particular, this goes in line with the results of\ \cite{Neves:2015ria} according to which no black hole
can form in an Ekpyrotic contracting phase where $w\gg 1$. However, one needs to be slightly careful in applying our results to a
model where the background is driven by a scalar
field\footnote{We conjecture that an oscillating scalar field with $c_\mathrm{s}^2=1$ would not lead to the formation of black holes.
Accordingly, the original matter bounce scenario would be stable against this type of instability. The situation is less obvious
for a scalar field with a non-canonical kinetic term in its action (e.g.,~a $k$-essence
scalar field), which could result in $c_\mathrm{s}^2\ll 1$.
In this case, the result might be closer to that of hydrodynamical pressureless matter where black holes are produced.}
since it may have $w\neq c_\mathrm{s}^2$, or equivalently, $w$ may be time dependent.

As we mentioned in the text, there remains to show that models of nonsingular bouncing cosmology which could have a mixture of matter and radiation
(e.g., the $\Lambda$CDM bounce and its extensions\ \cite{Cai:2014jla,Cai:2015vzv}) can still agree with observations.
To avoid the formation of black holes, radiation needs to dominate early enough, and in turn, this will
affect the perturbation modes that are of observational interest today and that
acquire a nearly scale-invariant power spectrum of curvature perturbations in the matter-dominated contracting phase.
In fact, it is known that the transition from matter domination to radiation domination would produce a break in the power spectrum
from scale invariance to a very blue spectrum. Such a break is highly constrained from observations,
and it implies that the radiation-dominated contracting phase must be shorter than in our expanding universe\ \cite{Li:2009cu}.
Yet, it appears to be still possible for these models to satisfy the observational constraints on the power spectrum
and avoid the formation of black holes in the contracting phase.

In this paper, we also showed that bouncing cosmologies that are solely driven by matter with $w=c_\mathrm{s}^2\ll 1$
(or for which the matter-dominated contracting phase lasts long enough before radiation dominates)
are not robust against the formation of black holes.
Since we find that these black holes form well before reaching the Planck scale,
the corresponding nonsingular bouncing cosmologies cannot ignore the formation of these black holes.
This agrees with the results of\ \cite{Banks:2002fe} which find an unstable growth of inhomogeneities
and the formation of black holes, hence the name ``black crunch'' that they gave to describe this scenario.

Finally, we showed that when the conditions for black hole formation are satisfied, the first black holes that form
are of Hubble size (the Schwarzschild radius is equal to the Hubble radius).
Once these Hubble-size black holes form, our perturbative analysis breaks down, hence we did not present the subsequent evolution of the
universe. Still, we can comment on a number of possible outcomes.

It is argued in\ \cite{Banks:2002fe} that such Hubble-size black holes behave as a $w=1$ fluid.
This leads to an alternative scenario to inflationary cosmology,
named holographic cosmology\ \cite{Banks:2001px,Banks:2003ta,Banks:2004vg,Banks:2004eb},
in which the so-called dense $p=\rho$ ``black hole gas'' serves as the seed to the observed large scale structure of our universe.
Also, in line with our motivation coming from bouncing cosmology, it is suggested in\ \cite{Veneziano:2003sz}
that such a dense black hole gas could lead to a model for the ``big bounce''.
The idea is that, in string theory, the black holes would evolve to become a dense gas of ``string holes'',
string states that lie along the correspondence curve between black holes and strings, as the string coupling evolves.
Furthermore, it is believed that the Hubble-size string holes saturate the conjectured cosmological entropy bound (see, e.g.,
\cite{Fischler:1998st,Veneziano:1999ts,Bousso:1999xy},
the review\ \cite{Bousso:2002ju} and references therein),
and thus, the entropy associated with the Hubble radius would be proportional to the area.
Since this is the same holographic scaling of the entropy and of the specific heat that is found in
string gas cosmology, one may hope to have a successful structure formation scenario just as in string gas cosmology
(see\ \cite{Brandenberger:1988aj,Nayeri:2005ck,Brandenberger:2006xi,Brandenberger:2006vv,Brandenberger:2006pr,Brandenberger:2014faa}
and also\ \cite{Brandenberger:2008nx,Brandenberger:2011et,Brandenberger:2015kga} for reviews).
Alternatively,\ \cite{Oshita:2016btk} proposes the idea
that a black hole could serve as a nucleation cite of a false vacuum bubble that could tunnel,
under some conditions and assumptions, to an inflationary universe, and thus, the black holes that naturally form
in a matter-dominated contracting universe could undergo such a tunneling and lead to inflationary universes.
At last, it could be that the black holes that are produced in the contracting universe simply ``pass through''
any given model of nonsingular bounce and form primordial black holes when they re-enter the Hubble radius,
as suggested by\ \cite{Carr:2011hv,Carr:2014eya}.
These would leave specific imprints in today's universe, in the form of, e.g., dark matter or
gravitational waves (see, e.g.,\ \cite{Clesse:2015wea,Clesse:2016vqa,Bird:2016dcv}), which could allow us to constrain the given model.

In summary, although the formation of black holes in a contracting universe
is an undesired feature in typical bouncing cosmologies, it seems to be of particular interest in many alternative scenarios
of the very early universe and may allow us to probe new physics and lead to the emergence of new ideas.
Consequently, we plan to expand upon the possible outcomes outlined above in more detail in a follow-up paper\ \cite{Quintin:2016}.

\paragraph*{Note added:}
While this paper was under preparation, we were informed that a similar study had been undertaken
by an independent group. This study reaches similar conclusions to ours with a slightly different approach\ \cite{Chen:2016kjx}.

\section*{Acknowledgments}

We thank Gabriele Veneziano, Tom Banks, and Jun'ichi Yokoyama for valuable discussions.
We also thank Yi-Fu Cai, Jie Wen Chen, Junyu Liu, and Hao-Lan Xu for making us aware of their work and for useful comments.
J.~Q.~acknowledges financial support from the Walter C.~Sumner Memorial Fellowship
and from the Vanier Canada Graduate Scholarship administered by
the Natural Sciences and Engineering Research Council of Canada (NSERC).
R.~B.~is supported by funds from NSERC and the Canada Research Chair program.

\end{document}